# Ordering transitions of weakly anisotropic hard rods in narrow slit-like pores

Roohollah Aliabadi[1),a], Péter Gurin[2)], Enrique Velasco[3)] and Szabolcs Varga[2)]

[1)] Department of Physics, College of Science, Fasa University, Fasa, Iran

[2)] Institute of Physics and Mechatronics, University of Pannonia, PO Box 158, Veszprém, H-8201 Hungary

[3)] Departamento de Física Teórica de la Materia Condensada, Instituto de Física de la Materia Condensada (IFIMAC) and Instituto de Ciencia de Materiales Nicolás Cabrera, Universidad Autónoma de Madrid, E-28049 Madrid, Spain

**Abstract**

The effect of strong confinement on the positional and orientational ordering is examined in a system of hard rectangular rods with length $L$ and diameter $D$ ($L>D$) using the Parsons-Lee modification of the second virial density- functional theory. The rods are nonmesogenic ($L/D<3$) and confined between two parallel hard walls, where the width of the pore ($H$) is chosen in such a way that both planar (particle's long axis parallel to the walls) and homeotropic (particle's long axis perpendicular to the walls) orderings are possible and a maximum of two layers are allowed to form in the pore. In the extreme confinement limit of $H \leq 2D$, where only one layer structures appear, we observe a structural transition from a planar to a homeotropic fluid layer with increasing density, which becomes sharper as $L \to H$. In wider pores ($2D<H<3D$) planar order with two layers, homeotropic order, and even combined bilayer structures (one layer is homeotropic, while the other is planar) can be stabilized at high densities. Moreover, first order phase transitions can be seen between different structures. One of them emerges between a monolayer and a bilayer with planar orders at relatively low packing fractions.

**Introduction**

It is still a fascinating topic from both fundamental and practical points of view to examine hard body fluids in slit-like pores as a model system of nanoparticles in nanopores [1,2]. The reason for this is that the connection between two- and three dimensional systems can be examined by tuning the wall separation ($H$) between the parallel walls. For example, although first order melting transitions of spherical hard particles weaken and become continuous with decreasing $H$, an intermediate hexatic phase emerges between fluid and solid structures as $H \to 0$ [3]. In addition, several crystalline structures can be generated such as the layered structures with square and triangle symmetry inside the pore. The stability of these phases depends on how many layers can accommodate between the confining walls [4,5]. The case of anisotropic hard particles is more complicated in pores as several mesophases can be present, which are stable even in the bulk limit such as the nematic, smectic A and columnar mesophases [6,7]. Smooth surfaces usually prefer the planar ordering at the walls and promote the formation of orientationally ordered phase [8], while rough ones may work against ordering [9]. With some surface treatments and rubbing it is possible to achieve either planar or homeotropic ordering at contact to the wall, which is very important in the development of display devices [10-12]. The ordering behaviour can be examined in hybrid cells, too, where one wall favours the planar ordering, while homeotropic alignment is induced at the other wall. The ordering of anisotropic mesogenic particles may be uniform, linear and even step-like due to the subtle interplay between the anchoring energies of the conflicting walls [13-17].

---

[a] Aliabadi@fasau.ac.ir or Aliabadi313@gmail.com

The nematic ordering of hard rods is very different in wide and narrow slit-like pores even if only excluded volume (steric) interactions are present. The wall-particle repulsion supports planar ordering to maximize the available space for the particles, i.e. the surface of the wall is always wet by a planar nematic film even if the system is very dilute and $H \to \infty$. However, the middle of the pore can be isotropic and a capillary nematisation (isotropic-nematic transition) may occur with increasing density. This capillary transition weakens with decreasing $H$ and terminates in a critical point [18,19]. The critical pore width ($H$) of the capillary nematisation is about 2-3 times larger than the length of the rod, but it depends weakly on the shape anisotropy of the particles [20]. Note that this is not a common property of all liquid crystals as the first order nature of the isotropic-nematic transition can survive even in the $H \to 0$ limit in some systems [21]. Another important phenomenon in the system of confined hard rods is the biaxial nematic wetting at the walls. Due to the strong adsorption of particles at the walls, an in-plane aligning transition can be induced with increasing density, which takes place between uniaxial nematic and biaxial nematic films [22,23]. This transition survives even in the extreme confinement limit $H \to 0$; therefore, it is interpreted as the isotropic-nematic transition of two-dimensional (2D) hard rods [24,25]. Along these lines, both experimental and theoretical studies have been devoted to the capillary nematisation and the surface ordering in strongly confined lyotropic and thermotropic liquid crystals [10, 26-30]. Even a second order isotropic-nematic phase transition can be observed in extremely confined semiflexible polymer solutions [31].

In our study we examine the effect of strong confinement on the orientational and positional ordering of non-mesogenic hard rods. This is motivated by our recent study of 2D hard rectangles, where the particles are weakly anisotropic and are confined between two parallel lines [32]. Using the exact transfer operator method it has been shown that planar-to-homeotropic orientational and monolayer-bilayer structural changes may arise in very strong confinements. These orientational and structural changes are continuous, without any nonanalytic behaviour in the thermodynamic quantities. However, the peak in the heat capacity and the sudden changes in the equation of state suggest that the confined rectangles undergo a strong structural transition which might turn to be a true phase transition in some three dimensional (3D) systems, such as the hard cuboids between two parallel walls. To our best knowledge only very few works are devoted to the ordering properties of weakly anisotropic 3D hard particles, where the width of the confinement is of the order of the particles dimensions. Khadilkar and Escobedo examined the positional ordering of polyhedral particles in slit-like pores, where the width of the pore is varied to see the structure of layered phases up to five layers [33]. Computer simulation studies of hard ellipsoids [34] and hard spherocylinders [35] support the belief that the strong confinement of weakly anisotropic particles promotes the formation of nematic order and layered structures even if the shape anisotropy of the particles is not enough large to form a nematic phase in bulk. Here we investigate the role of strong confinement on both orientational and positional ordering of hard cuboids and search for the possible phase transitions and structural changes at low and high densities.

**Theory**

We examine the orientational and positional ordering of rectangular hard rods, which are placed between two parallel hard walls with wall-to-wall distance $H$. The rod-like particles are modelled as rectangular cuboids with dimensions $D$ and $L$, where $D$ is the length of the two shorter sides, i.e. the cross section of the particle is a square, while $L$ is the length of the long side. The particle-particle and the particle-wall interactions are hard repulsive, i.e. the particles are not allowed to penetrate each other and to overlap with the confining walls. To determine the low and high density structures of the confined hard rods we resort to the classical Onsager's density functional theory with the Parsons-Lee modification [36,37] to make it suitable for weakly anisotropic particles. Even though this theory is approximate, it reproduces the simulation results for confined liquid crystals quite accurately [38]. Note that exact theoretical results can be obtained only in the quasi-one-dimensional limit [39] and in some lattice gases [40]. Here we present only the working equations and the differences from the equations of our previous study [20], where we studied the phase behaviour of long hard rods ($L/D>10$) in wide pores. As before, we restrict the orientational space of the long axis of the particle to three mutually orthogonal directions (Zwanzig approximation), which are chosen to be the Cartesian axes ($x$, $y$ and $z$), where $x$ and $y$ axes span the confining flat surfaces, while $z$ axis is perpendicular to the walls. We do not consider the possible in-plane positional order (crystallisation), which is known to be only quasi-long-ranged and the transition from isotropic fluid to solid is continuous Kosterlitz-Thouless type [41]. Therefore, the local density $(\rho)$ depends only on $z$ coordinate of the position vector $\vec{r}=(x,y,z)$ in our formalism. The packing fraction ($\eta$) with the above conditions can be determined from the local densities of orientations $x$, $y$ and $z$, i.e.

$$\eta = \frac{v_0}{V} \sum_{i=x,y,z} \int d\vec{r} \rho_i(\vec{r}) = \frac{v_0}{H} \sum_{i=x,y,z} \int dz \rho_i(z), \qquad (1)$$

where $v_0 = LD^2$ is the volume of the cuboid particle, $V = AH$ is the volume of the pore, $A$ is the area of the confining walls, $\rho_i(z)$ is the local density of the orientation $i$. The lower and upper limits of integrations in Eq. (1) are $(D/2, H-D/2)$ for $\rho_x$ and $\rho_y$, while $(L/2, H-L/2)$ for $\rho_z$. This means that the origin of the Cartesian system is in one of the walls. The key quantity to determine the local densities in inhomogeneous fluids is the grand potential ($\Omega$). On the level of Onsager's second virial theory with the Parsons-Lee modification, it is given by

$$\frac{\beta \Omega}{A} = \sum_{i=x,y,z} \int dz \rho_i(z) [\ln \rho_i(z) - 1 + \beta V_{ext}^i(z) - \beta \mu] + \frac{1}{2} c \sum_{i,j=x,y,z} \int dz_1 \rho_i(z_1) \int dz_2 \rho_j(z_2) A_{exc}^{ij}(z_1 - z_2), \qquad (2)$$

where $\beta = (k_B T)^{-1}$ is the inverse temperature, $\beta V_{ext}^i$ is the external potential acting on a particle with orientation $i$ and the walls, $\mu$ is the chemical potential, $c = (1-3\eta/4)(1-\eta)^{-2}$ is the Parsons-Lee prefactor and $A_{exc}^{ij}$ is the excluded area between two cuboids with orientations $i$ and $j$. The explicit expressions for $\beta V_{ext}^i$ and $A_{exc}^{ij}$ can be found in our previous work [20]. Note that Eq. (2) reduces to Onsager's theory in the low density limit ($\eta \to 0$), which corresponds to $c = 1$. In order to determine the equilibrium local densities, the grand potential has to be minimized with respect to all density components, i.e. $\delta(\beta \Omega / A) / \delta \rho_k(z) = 0$ ($k=x,y,z$). The functional differentiation results in

$$\ln \rho_k(z) + \beta V_{ext}^k(z) - \beta \mu + \frac{1}{2} \frac{dc}{d\eta} \frac{v_0}{H} \sum_{i,j=x,y,z} \int dz_1 \rho_i(z_1) \int dz_2 \rho_j(z_2) A_{exc}^{ij}(z_1 - z_2)$$
$$+ c \sum_{i=x,y,z} \int dz_1 \rho_i(z_1) A_{exc}^{ik}(z - z_1) = 0, \qquad (3)$$

where $\frac{dc}{d\eta} = \frac{5-3\eta}{4(1-\eta)^3}$ and we used that $\frac{\delta \eta}{\delta \rho_k(z)} = \frac{v_0}{H}$, which is coming from Eq. (1). After some rearrangements of Eq. (3), we can express the local densities as

$$\rho_k(z) = \exp\left[\beta\mu - \frac{1}{2}\frac{dc}{d\eta}\frac{v_0}{H}\sum_{i,j=x,y,z}\int dz_1 \rho_i(z_1)\int dz_2 \rho_j(z_2) A_{exc}^{ij}(z_1-z_2)\right] \times$$
$$\exp\left[-\beta V_{ext}^k(z) - c\sum_{i=x,y,z}\int dz_1 \rho_i(z_1) A_{exc}^{ik}(z-z_1)\right]. \quad (4)$$

Eq. (4) defines a self-consistent set of equations for $\rho_x$, $\rho_y$ and $\rho_z$, which can be solved iteratively at a given chemical potential, shape anisotropy and pore width. Note that $\frac{dc}{d\eta}=0$ in the original Onsager's formalism. Instead of working with $\beta\mu$, we substitute Eq. (4) into Eq. (1), which results in the following expression for the chemical potential

$$\exp[\beta\mu] = \frac{H\eta/v_0 \exp\left[\frac{1}{2}\frac{dc}{d\eta}\frac{v_0}{H}\sum_{i,j=x,y,z}\int dz_1 \rho_i(z_1)\int dz_2 \rho_j(z_2) A_{exc}^{ij}(z_1-z_2)\right]}{\sum_{l=x,y,z}\int dz_1 \exp\left[-\beta V_{ext}^l(z_1) - c\sum_{i=x,y,z}\int dz_2 \rho_i(z_2) A_{exc}^{il}(z_1-z_2)\right]}. \quad (5)$$

This expression allows us to replace $\beta\mu$ with $\eta$ in Eq. (4) and obtain a set of equations for the local densities in terms of packing fraction as follows:

$$\rho_k(z) = \frac{H\eta}{v_0} \frac{\exp\left[-\beta V_{ext}^k(z) - c\sum_{i=x,y,z}\int dz_1 \rho_i(z_1) A_{exc}^{ik}(z-z_1)\right]}{\sum_{l=x,y,z}\int dz_1 \exp\left[-\beta V_{ext}^l(z_1) - c\sum_{i=x,y,z}\int dz_2 \rho_i(z_2) A_{exc}^{il}(z_1-z_2)\right]} \quad (6)$$

We have solved iteratively the above coupled equations for the local densities at a given $\eta$, $L/D$ and $H$. The numerical integrations are performed with the trapezoidal quadrature. Having obtained the local densities from Eq. (6), we can calculate the grand potential and the chemical potential using Eqs. (2) and (5). First order phase transitions are located at the intersection point of two different solutions of Eq. (6) in the $\beta\mu - \beta\Omega/A$ plane, which corresponds to the equality of chemical potentials and that of pressures of the coexisting phases. In the following section we present our results in dimensionless units, where $D$ is the unit of distance, i.e. $\rho_i^* = \rho_i D^3$ and $H^* = H/D$.

**Results**

Even though our formalism can be applied for wide pores and long rods, here we consider only weakly anisotropic particles in very narrow pores and search for possible structural changes and phase transitions. The shape anisotropy of the cuboidal particle (*L/D*) has two restrictions: 1) *L/D*<3, i.e. the particles are nonmesogenic in bulk, and 2) *H*>*L*, i.e. the particles are allowed to stay both in planar (long axis of the particle is parallel to the walls) and homeotropic (long axis of the particle is perpendicular to the walls) orientations in the pore. We also restrict our attention to such narrow pores that a maximum of two fluid layers are allowed to form. This can occur only if *H*<3*D*. Before presenting the results, it is worth determining the possible one- and two-layer structures in the close packing limit to check the reliability of the forthcoming results. Three different close packing structures can be identified in our confined system with the above conditions, which can be summarized as follows: a) only one homeotropic layer is allowed to form for $H \leq 2D$ and for $2D < H \leq 3D$ if $L > 2D$, b) two layers with planar ordering for $2D < H \leq 3D$ if $L < 2D$ and $L > H - D$, and c) one layer is planar, while the other is homeotropic for $2D < H \leq 3D$ if *L*<2*D* and $L < H - D$. These close packing structures are shown together in Figure 1. Note that we do not consider those cases when two homeotropic layers can form in the pore, i.e. when *L*<*H*/2.

We first present our results for *H/D*=2, where only one layer of fluid is allowed to form both in planar and homeotropic orientations. In this case the local densities do not depend on *z* as all excluded areas are independent from $z_1 - z_2$ distance of the particles 1 and 2. The structure of the monolayer for varying *L/D* is shown in Fig. 2, where the fraction of particles in planar and homeotropic configurations are calculated from

$$X_P = \frac{\int dz [\rho_x(z) + \rho_y(z)]}{\sum_{i=x,y,z} \int dz \rho_i(z)} \text{ and } X_H = \frac{\int dz \rho_z(z)}{\sum_{i=x,y,z} \int dz \rho_i(z)}, \tag{7}$$

where the evaluation of the integrals are trivial for $H \leq 2D$ as $\int dz [\rho_x(z) + \rho_y(z)] = (H - D)(\rho_x + \rho_y)$ and $\int dz \rho_z(z) = (H - L)\rho_z$. Fig. 2 shows that the structure is dominantly planar at low densities, while it is homeotropic at high ones. As $L \to H$ the order is more and more planar for $\eta < 0.5$, i.e. $X_P \to 1$ and $X_H \to 0$. This can be attributed to the fact that the available space along $z$ axis in planar orientation is $H-D$, while the homeotropic ordering allows less positions for the particles as the available room along $z$ axis is just $H-L$. The reason why the monolayer becomes homeotropic for $\eta > 0.5$ is that the maximum packing fraction ($\eta_{max}$) is just 0.5 in planar, while it is $2D/H$ in homeotropic order. This means that the only way to form closely packed structure if the monolayer undergoes a structural change from planar to homeotropic order. One can show that $X_P = 2(H-D)/(3H-L-2D)$ at very low densities, because the translational and orientational entropy terms $(\rho_i \ln \rho_i)$ determines the structure of the system, while the packing entropy (excluded area) terms are negligible. From this expression one can see immediately that $X_P = 2/3$ for $L=D$ ($x$, $y$ and $z$ directions are equally probably) and $X_P = 1$ for $L=2D$ (all particles are in the $x$-$y$ plane). However, at high densities the packing entropy term wins over the translational and orientational ones because the excluded volume term can be minimized between the particles by homeotropic ordering ($V_{exc}^{zz} = \int dz A_{exc}^{zz} = D^2(H-L)$) in Eq. (2). Therefore the structural planar-homeotropic change is the result of a subtle competition between the translational, orientational and the excluded volume terms of the grand potential. Our search for possible in-plane orientational ordering has resulted into no biaxial order, i.e. $\rho_x = \rho_y$ for any $L < 2D$. This is not surprising because even 2D objects with a shape anisotropy less than 3 do not form a nematic phase in a plane.

In wider pores, i.e. $2D < H \leq 3D$, we encounter the situation that an inhomogeneous fluid is allowed to form with two layers and the close packing structure depends on $L$. We show the resulting phase diagrams in Fig. 3 for $H/D = 3$, 2.5 and 2.2 and the possible structures in Fig. 4. These are the followings: uniaxial planar (UP), biaxial planar (BP), homeotropic monolayer (H), and special bilayers with homeotropic order at one end and uniaxial or biaxial planar order at the other one (UT and BT). The previously discussed planar-homeotropic structural change survives at $H/D = 3$ if $L/D > 2$, but it is now first order transition between planar bilayer and homeotropic monolayer phases. The coexisting planar phase can be either uniaxial ($\rho_x = \rho_y$) or biaxial ($\rho_x \neq \rho_y$) as the second order uniaxial-biaxial (UP-BP) transition crosses the biphasic region. Therefore the following phase sequences can be seen with increasing density for $H/D = 3$ (Fig. 3(a)): (1) UP → BP → H for $2.5 < L/D < 3$ and $2 < L/D < 2.1$ and

(2) UP $\to$ H for $2.1 < L/D < 2.5$. The density profiles of these phases indicate that: (i) UP phase consists of two fluid layers with strong adsorption at the walls and a few percent of the particles with homeotropic order in the middle of the pore (Fig. 4(a)); (ii) in BP phase, particles mostly align along the x axis at both walls (Fig. 4(b)); and (iii) in H phase, particles order along the normal of the confining plates without adsorption at the wall (Fig. 4(c)). It is clear that the UP-BP transition line moves up in density with decreasing shape anisotropy as the excluded area gain becomes less with in-plane ordering. The widening stability region of the BP phase for L/D>2.5 is due to the translational entropy loss along the z axis in the homeotropic phase as $L/D \to 3$. However, it is harder to understand the stabilization of the planar phase with respect to homeotropic order for L/D<2.5 because the translational entropy term is now larger in the hometropic order. Therefore, the packing entropy may be the main factor in this phenomenon as the close packing planar and homeotropic densities are becoming closer to each other as $L/D \to 2$. The existence of biaxial structures at lower L/D is due to the stronger destabilization of the homeotropic ordering than that of UP-BP transition as $L/D \to 2$. The situation is dramatically changed for $L/D < 2$ (see Fig. 3(a)), because combined bilayer phases (UT and BT) can be more packed than the planar and homeotropic structures. The density profiles for the case $L/D = 1.85$ with increasing density show that the planar particles are depleted from the middle of the pore at $\eta = 0.3$ (Fig. 4. (d)), both planar and homeotropic particles are adsorbed at the walls and the central region of the pore becomes practically empty at $\eta = 0.55$ (Fig. 4. (e)), segregation of the homeotropic and planar particles occurs at the opposite walls at $\eta = 0.65$ (Fig. 4. (f)) and even biaxial order can take place in the planar layer at $\eta = 0.75$ (Fig. 4. (g)). Interestingly, the transition between UP and UT phases is of first order and becoming weaker with decreasing $L/D$. The disappearance of this transition is due to the fact that a homeotropic bilayer structure emerges for $L/D < 1.5$, which can be more packed than the combined one. Biaxial order can only be stable at very high densities through a second order UT-BT transition, as the shape anisotropy is very weak. The phase diagram in narrower pores ($H/D = 2.5$ and $H/D = 2.2$) is similar, but the coexistence curve between planar bilayer and homeotropic monolayer structures is shifted to higher packing fractions and the coexisting planar bilayer is biaxial as there is no intersection between the first and second order transitions (see Fig. 3(b)). However, the region of $H - D < L < 2D$, which is not empty for $H < 3D$, is special in the sense that the close packing can be achieved with the planar bilayer structure. Therefore it is interesting that planar bilayer ordering develops continuously with increasing density for $H/D = 2.5$ in the range of $1.5 < L/D < 2$. In narrower pores ($H/D < 2.3$) we have found that the process of planar bilayer ordering is accompanied by a first order transition between monolayer and bilayer phases (see Fig. 5). The density distributions of these structures are shown in Figs. 4(h) and (i). We can see that the hard rods do not form a bilayer up to the transition point with increasing density, because the density profiles are uniform with some adsorption at the walls in planar orientation, $\rho_x = \rho_y > 0$ in the middle of the pore, and with lots of particles in homeotropic orientation (see Fig. 4(h)). In addition, there is no reason to form a bilayer as the packing fraction is below the maximal packing fraction of the planar monolayer structure $(\eta_{max}(\text{monolayer}) = D/H)$. Fig. 4(i) shows that the confined fluid consists of two planar layers at higher densities, as $\rho_x = \rho_y \approx 0$ in the middle of the pore, and density profiles are strongly peaked at the walls, with $\eta > \eta_{max}(\text{monolayer})$. It can also be seen that planar bilayers are interrupted by homeotropic particles, as $\rho_z > 0$ in the middle of the pore. Therefore the phase transition in Fig. 5 can be considered as a monolayer-bilayer layering transition, which has been observed so far in a system of confined hard rods with larger shape anisotropies and with homeotropically aligning walls [42]. The existence of biphasic islands can be attributed to the presence of homeotropic rods. The left end of the island is due to the weak shape anisotropy as homeotropic short rods can easily reorient into the planar direction. The right end of the biphasic region can be also understood because there are less and less homeotropic rods with increasing shape anisotropy as translational entropy decreases. This can be seen clearly in the shift of the right end towards lower values of *L/D* with decreasing pore width *H*. The width of the pore is also an important factor in the stabilization of first-order layering transitions, because a substantial loss in the translational entropy is necessary during the formation of the bilayer. This happens when two layers cannot accommodate easily into the pore, i.e. $H \approx 2D$. This is the reason why we do not observe layering transitions for $H/D \geq 2.3$. Note that the biphasic island exists for $H/D < 2.1$, but that it moves into the direction of very high packing fractions and becomes smaller.

**Conclusions**

We have shown that a system of weakly anisotropic rod-like particles can undergo a phase transition in narrow slit-like pores where the particle-particle and particle-wall interactions are hard repulsive. Therefore the system is athermal and entropy driven. The systems investigated are practically quasi-two-dimensional as the pore width is chosen such that a maximum of two layers are allowed to form between the confining hard walls. Depending on the shape anisotropy of the rods, pore width and density, several mesophases can be realized, such as orientationally ordered monolayers and bilayers with in-plane uniaxial or biaxial order. Although the hard walls always support the planar ordering to maximize the translational entropy, the packing entropy may win over the other entropy terms and give rise to homeotropic ordering at high densities. If only a monolayer is allowed to form, i.e. $H<2D$, the wall-induced planar monolayer transforms continuously into a packing-entropy-induced homeotropic monolayer with increasing density. In wider pores $2D<H<3D$, as a result of the competition between translational, orientational and packing entropy terms, our scaled Onsager theory predicts first order phase transitions between monolayers and bilayers. The low density planar monolayer may transform continuously or discontinuously into middle density planar bilayer. At higher densities the bilayer may exhibit a second order ordering transition from uniaxial to biaxial phases, which takes place in the plane of the walls. In addition to this, the planar bilayer may transform into homeotropic monolayer or UT bilayer through first order phase transitions. Among these transitions, the monolayer-bilayer layering transition of the cases $H/D<2.3$ takes place at the lowest packing fractions $(\eta \sim 0.5)$, which is far from the maximal $\eta$ of the planar bilayer $(\eta_{max} = 2D/H \sim 1)$. Interestingly the UP-UT transition can take place at similarly low packing fractions if $2L \sim H$. However, the UP-H transitions are located at very high densities, which are sometimes very close to the close packing of the homeotropic order, i.e. $\eta \sim \eta_{max} = L/H$. Therefore, it may happen that the confined bilayer fluid freezes first, and then the quasi-two-dimensional crystal transforms into the homeotropic one, both coexisting phases being crystalline. To resolve this issue, an extension of the theory for in-plane positional ordering is needed. Finally, we mention that our previous 2D exact results are consistent with our present mean field results in the sense that confined hard rectangles exhibit very similar structural changes [32].

It is remarkable that a planar-to-homeotropic transition is observed in the presence of flat surfaces with increasing density. Such a phenomenon is also observed in a system of hard Gaussian overlap particles with a special wall-needle interaction [14] and in stiff ring polymers on hard walls [43]. Other examples belong to the realm of thermotropic liquid crystals, where the surfaces of the cell are treated and the planar-to-homeotropic transition takes place with decreasing temperature [44]. It remains an open question whether the density induced planar-to-homeotropic ordering survives in wider pores and whether a bistable device or a pressure sensor could be fabricated with such hard particles [45].

## Acknowledgements

P. G. and S. V. would like to thank the financial support of OTKA (Hungary) under Grant No K124353. E. V. is partially supported by MINECO (Spain), under Grant FIS2013-47350-C5-1-R. R. A. acknowledges Fasa University for supporting financially the research and for its computing facilities.

**Figure caption**

**Figure 1**

Cartoons show the close packing structures of hard rectangular rods with square cross sections in a slit-like pore. A single homeotropic layer (upper panel), a planar bilayer (middle panel) and a combined bilayer (lower panel) structures can be realised with the shown $H/D$ and $L/D$ parameters.

**Figure 2**

Planar-to-homeotropic structural change of hard rods in a narrow slit-like pore ($H/D$=2), where only one layer of particles is allowed to form. Fraction of hard rods in planar ($X_P$) and homeotropic ($X_H$) ordering is shown as a function of packing fraction for $L/D$=1.9, 1.99 and 1.999.

**Figure 3**

Phase diagram of confined hard rods in the packing fraction – shape anisotropy ($\eta$–$L/D$) plane for wall-to-wall separations of $H/D=3$ (upper panel), $H/D=2.5$ (lower panel) and $H/D=2.2$ (inset of the lower panel). The curves show the boundaries of the different structures. The following structures are found: uniaxial planar (UP), biaxial planar (BP), homeotropic (H), uniaxial (UT) and biaxial combined two-layer (BT) structures. The symbols designate few stable points of the observed phases. The structures of the hard rods at these symbols are shown in Figure 4. The biphasic regions are shaded with grey colour.

**Figure 4**

Density profiles of the stable structures as a function of $z^*=z/D$ in wide ($H/D=3$) and narrow ($H/D=2.25$) pores at different packing fractions and shape anisotropies, which can be identified with the help of the shown symbols in Figs. 3 and 5. The local density components $\rho_x^*(z)$ (black), $\rho_y^*(z)$ (red) and $\rho_z^*(z)$ (blue) are shown together in the diagrams. The following stable structures are seen: a) uniaxial planar $\left(\rho_x^*(z)=\rho_y^*(z)>\rho_z^*(z)\right)$, b) biaxial planar $\left(\rho_x^*(z)\neq\rho_y^*(z)\neq\rho_z^*(z)\right)$, c) homeotropic $\left(\rho_x^*(z)=\rho_y^*(z)\ll\rho_z^*(z)\right)$, d) uniaxial weakly planar with non-negligible portion of particles in homeotropic order, e) uniaxial strongly planar with some homeotropically ordered particles at the vicinity of the walls, f) uniaxial bilayer with homeotropic order at the left wall and planar order at the right wall, g) biaxial bilayer, h) uniform monolayer with weak adsorption in planar order at the walls and i) uniaxial planar bilayer with few particles in the homeotropic order.

**Figure 5**

Phase coexistence between monolayer and bilayer fluids in packing fraction – shape anisotropy ($\eta$–$L/D$) plane for pore widths $H/D=2.1, 2.15, 2.2$ and $2.25$. The density profiles of the uniform and planar phases, which are marked by symbols, can be seen in Figs. 4 h) and 4 i).

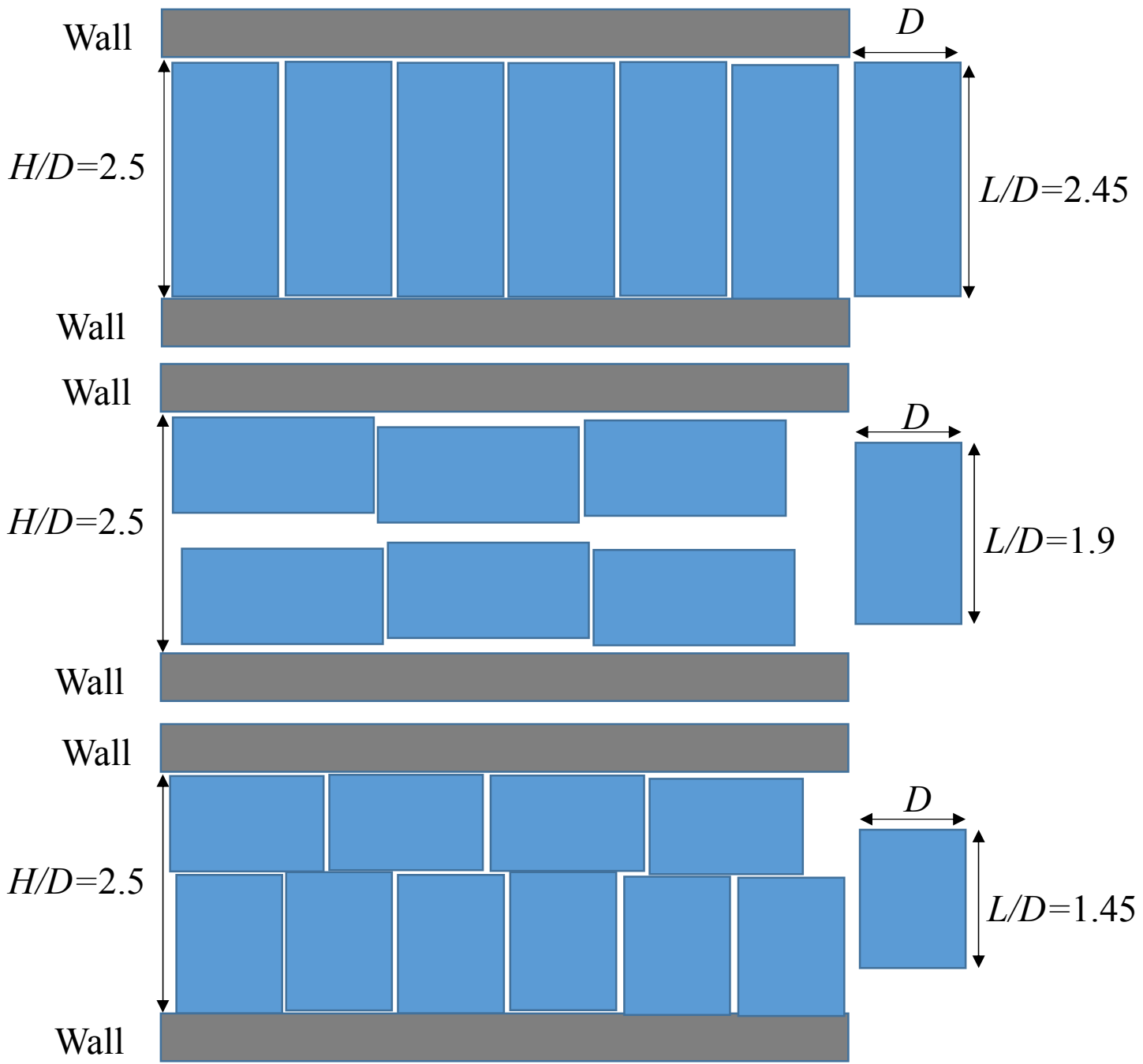

**Figure 1**

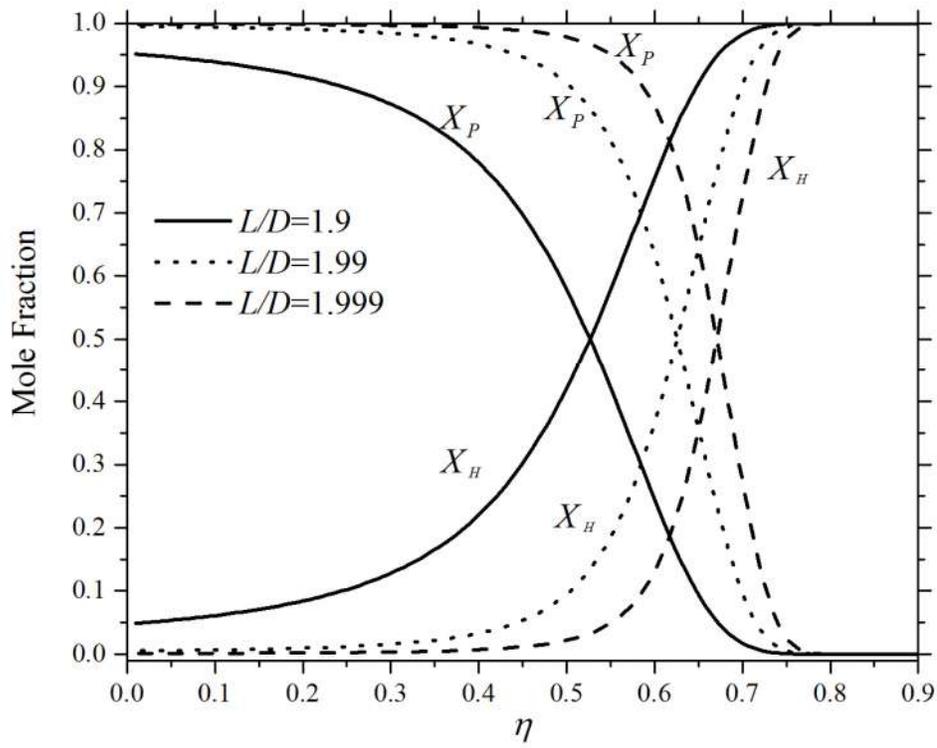

**Figure 2.**

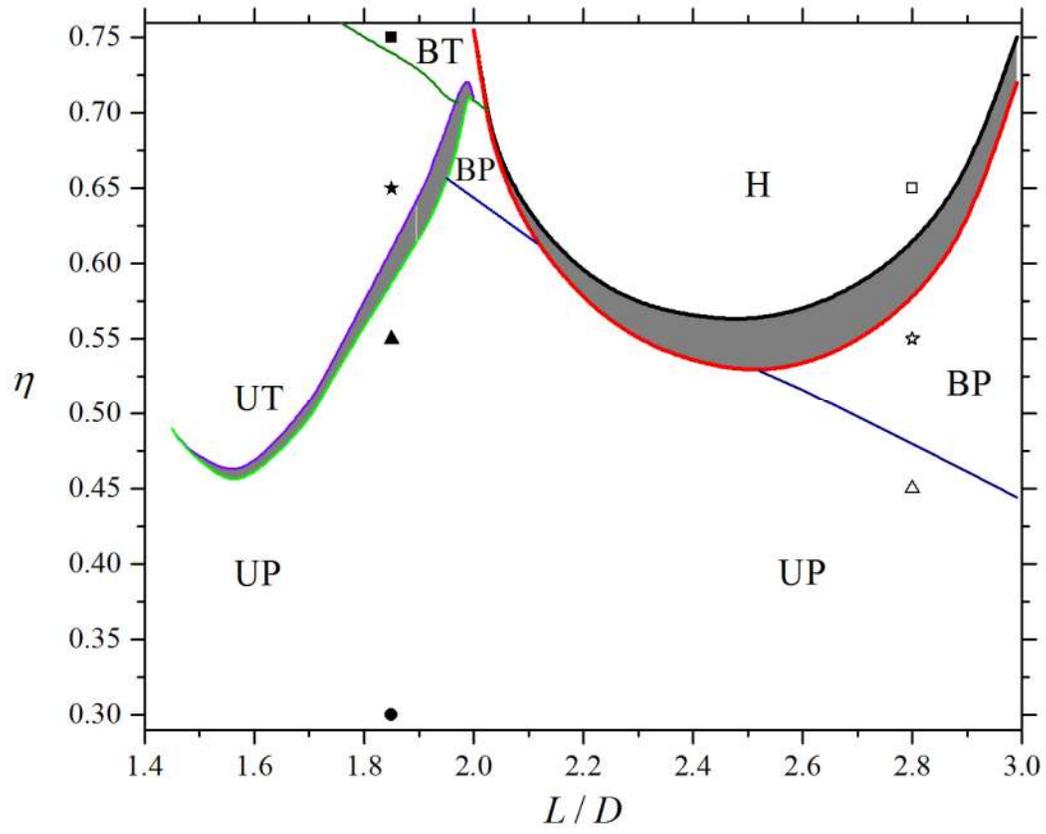

**Figure 3a**

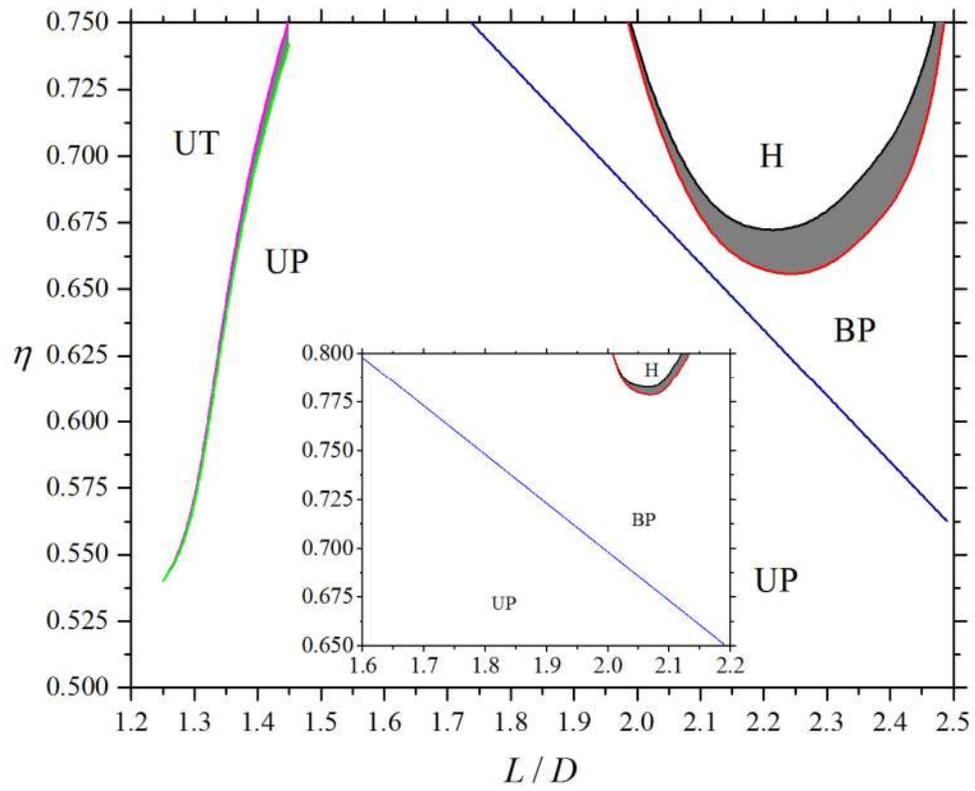

**Figure 3b**

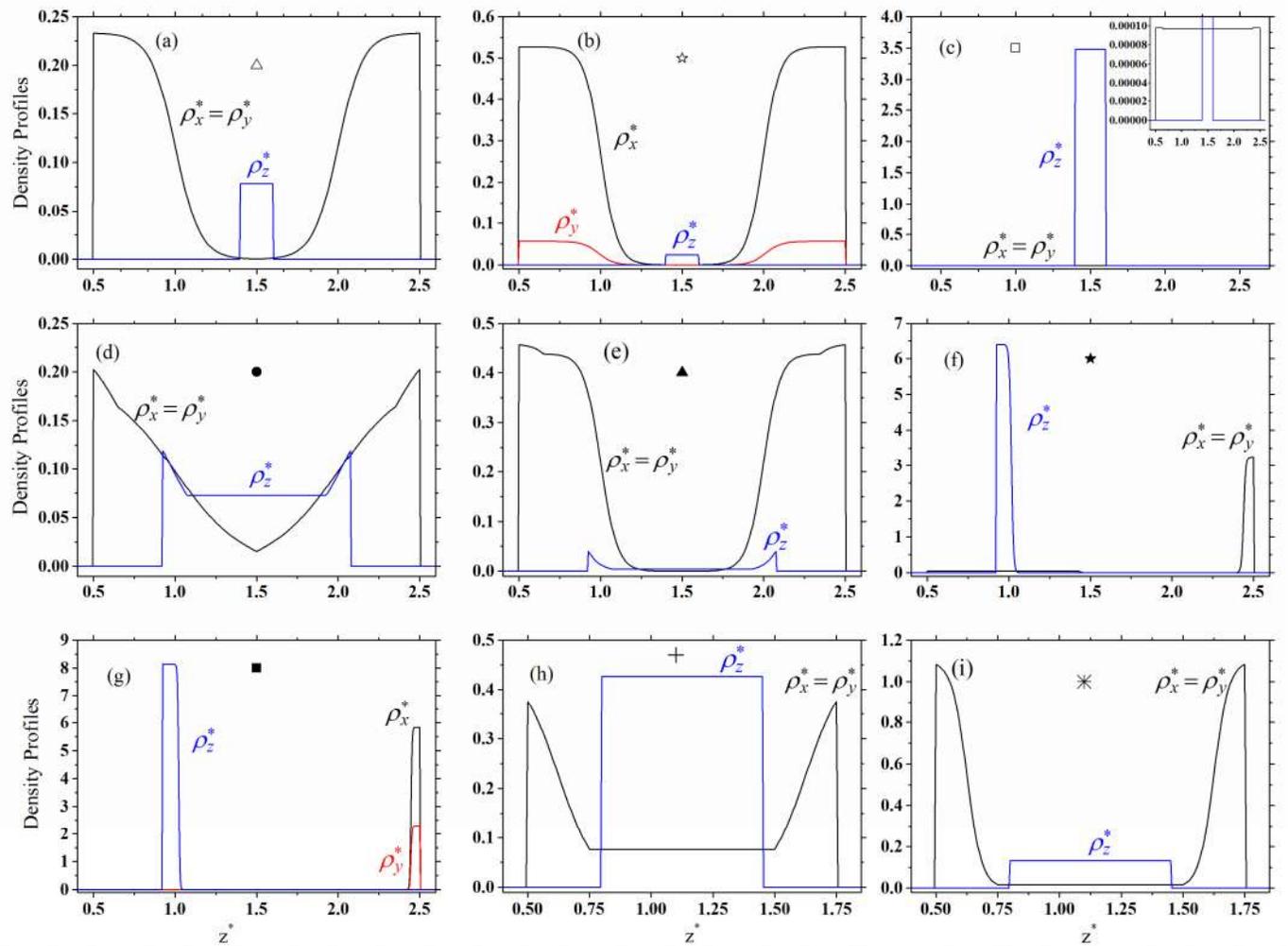

**Figure 4.**

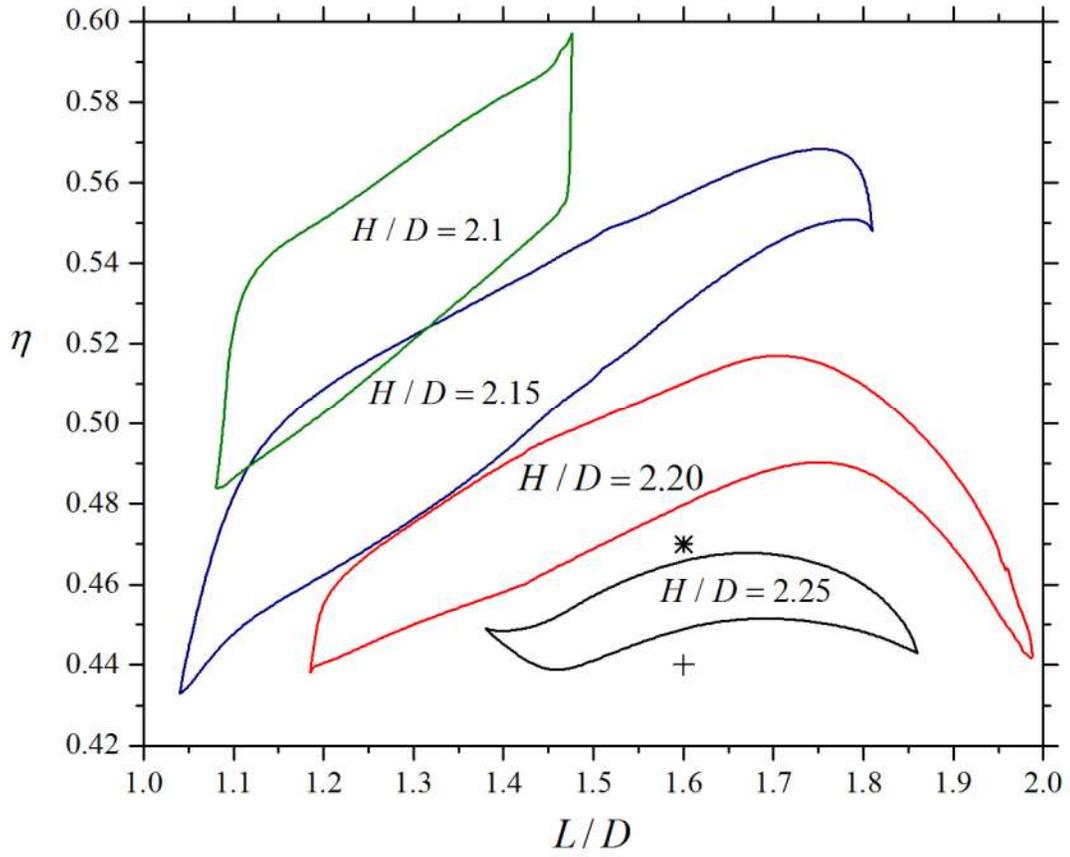

**Figure 5.**